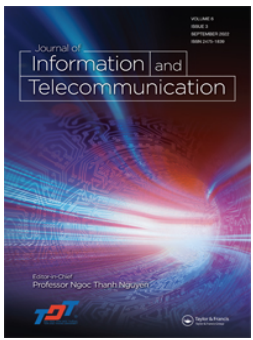

# A customer satisfaction centric food delivery system based on blockchain and smart contract

A. A. Talha Talukder, Md. Anisul Islam Mahmud, Arbiya Sultana, Tahmid Hasan Pranto, AKM Bahalul Haque & Rashedur M. Rahman

Submit your article to this journal

View related articles

View Crossmark data

# A customer satisfaction centric food delivery system based on blockchain and smart contract

A. A. Talha Talukder[a], Md. Anisul Islam Mahmud[a], Arbiya Sultana[a], Tahmid Hasan Pranto[a], AKM Bahalul Haque[b] and Rashedur M. Rahman [a]

[a]Department of Electrical and Computer Engineering, North South University, Dhaka, Bangladesh; [b]Software Engineering, LENS, LUT University, Lappeenranta, Finland

**ABSTRACT**

Food delivery systems are gaining popularity recently due to the expansion of internet connectivity and for the increasing availability of devices. The growing popularity of such systems has raised concerns regarding (i) Information security, (ii) Business to business (B2B) deep discounting race, and (iii) Strict policy enforcement. Sensitive personal data and financial information of the users must be safeguarded. Additionally, in pursuit of gaining profit, the restaurants tend to offer deep discounts resulting in a higher volume of orders than usual. Therefore, the restaurants and the delivery persons fail to maintain the delivery time and often impair the food quality. In this paper, we have proposed a blockchain and smart contract-based food delivery system to address these issues. The main goal is to remove commission schemes and decrease service delays caused by a high volume of orders. The protocols have been deployed and tested on the Ethereum test network. The simulation manifests a successful implementation of our desired system; with the payment being controlled by our system. The actors (restaurant, delivery-person or consumer) are bound to be compliant with the policies or penalized otherwise.

## 1. Introduction

The food sector is continually evolving with new technologies and business strategies. The online food delivery system has become a new trend catching popularity. With the fast-paced lifestyles, there appears to be little time for cooking or even sitting down to a three-course dinner at a restaurant, unless it is for a special occasion. Individuals order online using online applications to multitask and save time and money. Due to the fact that the third-party applications lack the capability to establish trust among the participants in a food-delivery ecosystem, it created several complications for customers. Either the customer receives their food cooked earlier and therefore not warm during the delivery. It is due to the violation of time constraints or the customer may not receive any food after placing an order. Particularly, forced exertion of policies are completely missing in third-world countries. On the other hand, customer satisfaction is critical to the escalating online food delivery sector. Food delivery app organizations, which establish a link between the customer and the restaurant via their app and deliveryman now determine the price of food and profit from each transaction. They provide substantial discounts on foods to attract new customers, which leads to considerable losses for smaller restaurants (Niharika, 2020). Discounting and lucrative advertisements bring in more customers. To triumph this race and win over the market, the restaurants take as much order as they can and often fail to deliver within time. Moreover, rides working on commission per delivery often mimic the same to gain more money. In both cases, customers are the sufferers.

Despite being a topic of importance, there exists a lack of focus on their concurrent impact on customer online loyalty, particularly in the context of online food delivery services in comparatively less developed countries (Suhartanto et al., 2018). To address these issues mentioned above in the online food delivery system, this study will reduce discounting race, issues raised by commission-based services, and focus on increasing customer satisfaction and loyalty over online food delivery services. More precisely, this study seeks to determine:

(1) the direct effect of e-service and food quality on online loyalty; and
(2) the indirect effect of perceived value and customer satisfaction via the mediation role.

The current strategies in online food delivery services lack customer convenience. Solving these issues within the food delivery system is critical for ensuring customer loyalty over these platforms (Suhartanto et al., 2018). Blockchain technology (Nakamoto, 2018), when combined with smart contracts (Szabo, 1997), has the potential to alleviate these challenges by creating a distributed network of restaurants, customers, and delivery persons. An organization or appropriate body will establish food services and pricing in a distributed de-centralized application. Using smart contracts, the system will ensure strict application of the regulation. The contract will then be stored in the Blockchain, which will be impenetrable to alteration or tampering. The commercial terms and conditions that govern online food delivery transactions become set in stone and unchangeable. The paper's primary contribution is summarized below.

- We propose penalty-based order processing in a decentralized food delivery ecosystem. Penalization in terms of money forces the actors to be compliant with the regulation.
- We propose a reputation mechanism in accordance with the penalization for increasing trust among the actors.
- We ensure data security throughout the system using blockchain and smart-contract based decentralized implementation.
- We propose a service that replaces the commission-based service in the online food delivery sector and increases customer satisfaction.

The remainder of the paper is organized as follows. The following section (section 2) comprises a detailed discussion of the relevant literature. Section 3 provides an overview of the proposed work mentioning the system components, actors and their role within the system. Section 4 describes the implementation details and evaluation process of our system. Section 5 contains the testing of our system from order placing to delivery - simulating a complete run of the system and at the same time discusses the advantages and challanges. The document ends with the concluding remarks in the last section.

## 2. Literature review

Blockchain and smart contract are two technologies that go hand-in-hand to build systems that need openness and rigorous proof of documentation. On par with the Blockchain, which establishes and maintains trust between participants, smart contracts operate independently of any third party or human being to maintain that trust inside the blockchain network. This section will briefly outline the background of blockchain technology, smart contracts, and their possibilities to overcome the current flaws of online food delivery services.

### 2.1. Blockchain

Blockchain technology has brought versatility to the fourth industrial revolution (Bodkhe et al., 2020). This technology gained its much-acclaimed recognition after developing cryptocurrencies such as Bitcoin to perform peer-to-peer transactions without a third-party involvement (Nakamoto, 2018). A blockchain is a distributed, immutable, and previous block, timestamp of the block creation, nonce and transactions represented in append-only data structure comprising a chronologically and cryptographically linked series of blocks (Wood Daniel Davis, 2014). Each block contains the pointer (hash) to the form of Merkel tree (Bodkhe et al., 2020). Blockchain technology can identify, record, validate, and process all transactions, facilitating the massive flow of information and securely updating in the network (Wesley, 2017). The structure of a blockchain network is delineated in Figure 1.

Blockchain technology significantly enhances the security of mobile cloud data, and it is highly beneficial for the upcoming generation (Hassija et al., 2021). Farouk et al. showed how Blockchain could impact while dealing with sophisticated medical data via IoT in terms of regulation, privacy preservation and information security while also aiding secure sharing and analysis of these data (Farouk et al., 2020). The rate of file loss in existing cloud storage systems can be as high as 100%, but the rate of file loss in the Blockchain is nearly zero (Yang et al., 2019). There is no integrated solution for processing distributed data autonomously in a blockchain-based context. This is possible through the use of smart contracts and blockchain technology. Numerous applications rely on blockchain technology as a foundation because to its inherent resistance to alteration. In the food business, the blockchain is used to ensure food safety (Tse et al., 2017). Ngamsuriyaroj et al., (2018) have presented a secure package delivery method based on Blockchain technology that protects both data and users' integrity and confidentiality. The proposed technique was assessed in terms of inserting and reading/writing data. In terms of security, Blockchain is on par with cloud storage options on the market (Hasan et al., 2020). Blockchain is by far one of the most secure ways for e-transaction than any other media. Roeck et al. (2019) shows that Blockchain's immutable, secure, and consensus-based ledger technology possesses the elements needed to reduce the effect of third parties in a supply chain, and they also provide evidence of Blockchain being cost-effective to supply chains in the long run.

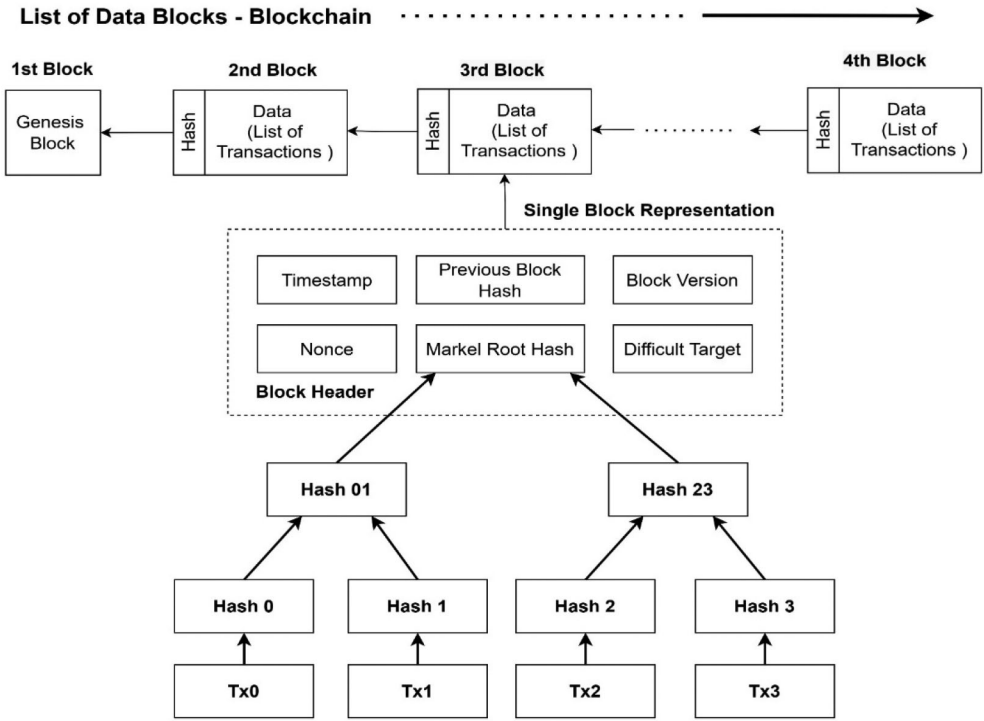

**Figure 1.** Blockchain structure.

## 2.2. Smart contract

A smart contract is a pre-written computer program that enables digital transactions under particular conditions or terms. Without requiring human interaction, smart contracts enable the tracking and implementation of complex agreements between parties (Novo, 2018). Smart contracts deployed inside a blockchain-based platform can assist with the self-execution of business logic to self-execute in a buyer-seller transaction network (DeCusatis et al., 2018; Valenta & Sandner, 2017). Each node in the network has its copy of the smart contract that operates autonomously and independently based on the triggering transaction data (Stanciu, 2017). Figure 2 shows the steps of a smart contract formulation. Encapsulating and protecting information and keeping it simple throughout the network is the primary goal of smart contracts. Smart contracts are used in various use cases; for example, a smart contract may represent an item's shipping cost, which fluctuates depending on arrival time constraints. According to the terms agreed upon by both parties and documented in the ledger, monies shift automatically when an item is received (Yewale, 2018). The critical want for establishing faith between two individuals may be met via smart contracts (Bader et al., 2019). Hawk is a framework for developing smart contracts with privacy-preserving features. Hawk compiler is responsible for compiling the software into a cryptographic protocol used to communicate between the Blockchain and the users (Kosba et al., 2016).

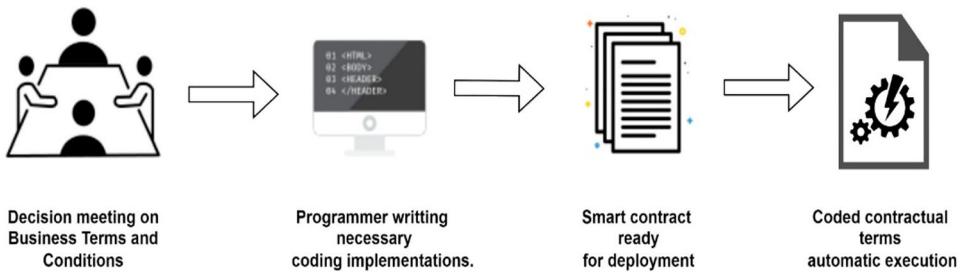

**Figure 2.** Steps of building a typical smart contract.

BMC Protector is a music copyright management system built on top of blockchain and smart contracts. Their smart contract manages all of the system's critical tasks, from composition through royalty distribution. It is exceedingly difficult to compromise and alter a smart contract that has been distributed inside the blockchain environment, which justifies ultimate security when implementing a smart contract within the blockchain ecosystem (Bader et al., 2019; Zhao & O'Mahony, 2018). With B-Ride, drivers may offer ride-sharing on their own, without going to any third-party applications they can provide their services. It is up to both customers and drivers to decide if they want to share a ride while preserving their data. As a result, fraudulent individuals might go incognito, which Blockchain provides to submit several requests to get a better deal (Baza et al., 2019). To create a block on a blockchain, there must be a common agreement. Smart contracts make it possible for any real-world application to insert business logic into the process on which the participating actors can relentlessly rely upon. When a block is created from any connecting node, the block creation information is broadcasted to every node within the network. Smart contract verifies the data in the new block against the agreement and, if it is legitimate, and meets the other criteria, then the block is added to the Blockchain. Following that, the blockchains of each network participant are updated.

In most cases, smart contracts are used to exert business logic to be checked before producing a block on the Blockchain (Yewale, 2018). When constructing a smart contract, it is common to utilize terms such as attributes, functions, modifiers, and events. Within a system, attributes represent storage variables in which values can be stored and modified. A function represents a chain of commands or processes. When a function is invoked, it performs the task indicated in the function's body. Events and modifiers are the next two in line where events enable the blockchain transaction log to store virtually anything. After occurring an event, it creates data. The data then goes directly to transaction logs to preserve the historical data to be retrieved later. This event triggering enables the structure to be auditable. Modifiers allow the modification behaviour of the functions in smart contracts. It has a range of applications, including limiting who can unlock functions after a specified time, perform a specific function, and so on (Pranto et al., 2021). Once the agreement's requirements are satisfied, smart contracts are immediately executed. This eliminates the need for a third party such as certain apps (Foodpanda, HungryNaki, UberEats), banks etc. The smart contract uses its agreement to control the system like a business, and the Blockchain gives a safe platform for storing and preserving data.

### 2.3. The present centralized system and its issues

(1) **The absence of rules governing deep discounting:** Food delivery app operators and aggregators entice users with significant discounts (Granheim et al., 2020). Conversely, restaurant owners and food industry owners face the brunt of extreme discounts. There are no clear criteria for regulating such situations, which results in lesser earnings for both sides.

(2) **Delayed food delivery:** Customer satisfaction is decreased when food is delivered late (Worku & Legoabe, 2017). As a result, external factors such as traffic, order volume, and weather conditions must be considered. Occasionally, a lack of training, a staffing deficit, inappropriate scheduling, or inefficient route planning might result in an improperly managed delivery routine. Some customers want foods from different places, which produces a routing and timing problem (Kohar & Jakhar, 2021). Additionally, it requires considerable work on the delivery platform's customer care employees to placate clients in such instances.

(3) **Chaotic order:** For restaurant proprietors, keeping track of internet demand and available resources becomes challenging. As a result, they receive a high number of orders and frequently cannot fulfil them. Deliverymen rush to the restaurant because of the multiple orders and waste or are forced to deliver more than one at a time (Liao et al., 2020). As a result, deliveries are delayed, and food quality deteriorates.

(4) **A scarcity of safe payment methods:** Customers will avoid ordering from food delivery platforms that do not accept various payment methods. They anticipate safe payment options and an efficient return process. Three significant causes have contributed to the growth of food delivery services. One is generation Y's use of digital technologies such as mobile phone applications. Second, the rise of new industry leaders and third-party applications such as Uber Eats and FoodPanda. They are market leaders, utilizing their centralized platform to manage restaurants, deliveryman, and customers. Occasionally, deliverymen receive minimal commissions due to their schemes, which causes them to violate company laws and regulations, which ultimately harms customers because they are the ones who need food on time. Third-party applications also have difficulties with time since they can adjust it based on the deliveryman's position, which varies over time and provides an inaccurate approximation. This forces the deliveryman to accept many orders concurrently, putting the customer in jeopardy once more. The same is true for restaurants, which accept numerous orders at a time because third-party apps do not have a punishment system for being late or failing to deliver food on time. As a result, the restaurant is forced to accept a large volume of orders and fails to deliver the food on time to the deliveryman, resulting in an unhappy customer. The major players generate money by charging for platform usage and managing their drivers as independent contractors. Thus, the benefits of blockchain technology may include the ability to transition from a B2B2C business model to a B2C one by connecting customers directly to restaurants and eliminating service costs associated with money redistribution. Quality control and continuous improvement are critical in online food delivery. The application of technology such as Blockchain and smart contracts will improve the existing state of food delivery. These enhancements are necessary to maintain a traceable, rigorous, and secure delivery procedure.

### 2.4. Blockchain integrated food delivery

Blockchain-based technologies are poised to reshape the food delivery sector in ways never seen before. Thus, the days of people being content just because they received a free pizza delivered within 30 minutes are over. Customers in the modern-day want convenience and quickness. Due to digitization, food delivery services have risen substantially in the last five years. One of the most recent technologies accessible is 'Wooberly Eats,' which mixes machine learning and blockchain technology. Wooberly utilizes 'Flutter,' an open-source UI framework, to provide a variety of capabilities, including data exchange with the Customer, Driver, Restaurant, and system administrator(s).

The growing popularity of smartphones and the convenience of purchasing via mobile applications quickly made them a customer favourite. As a result, an increasing number of individuals are growing used to ordering foods online for immediate or scheduled delivery (Liu, 2019). Online food delivery businesses are classified into two types (Yeo et al., 2017). The first is a restaurant-to-customer delivery business, such as Kentucky Fried Chicken or McDonald's, which may offer online food delivery services directly or through third-party operations. The second is a platform for a platform to customer delivery services such as Uber Eats, Foodpanda, and Hungrynaki. Online food delivery systems combine a diverse range of partner restaurants and provide delivery services to them. Modern life is made better by online food delivery systems. Customers may choose from a diverse selection of restaurants and cuisines to enjoy delectable meals (Chen et al., 2021). As new business models for app-based food delivery emerged, further obstacles became apparent (Lau & Ng, 2019).

The food delivery apps do not maintain time, leading to chaotic orders and dissatisfied customers; thus, the restaurant's reputation drops. Deliverymen also use this time manipulation opportunity in hand and take as many foods as possible for multiple deliveries at a time and this causes them to be late for delivering foods. So, again dissatisfied customers and drop of restaurant reputation. In this paper, we are solving this time hindrance issue and a chaotic order acceptance by restaurants by using Blockchain and smart contracts, which will also be a safe transaction method.

## 3. Proposed blockchain-based model

### 3.1. System overview

Our primary goal is to demolish the current centralized food distribution system, and our primary recommended solution will be decentralized. More precisely, we want to offer a blockchain-based food delivery system. The problems with the present system have been dealt with, and we want to create a win-win situation for everyone involved, including the customer, the restaurant, and the delivery guy. The general system overview is shown in Figure 3. We present an online food ordering and delivery administration platform that any entrepreneur can utilize to start their own food delivery business. The system is built on Blockchain technology to address various legitimate concerns that existing online food delivery stages face. It enables transactions via cryptocurrency, which provides decentralization and security, as discussed in the research discussion.

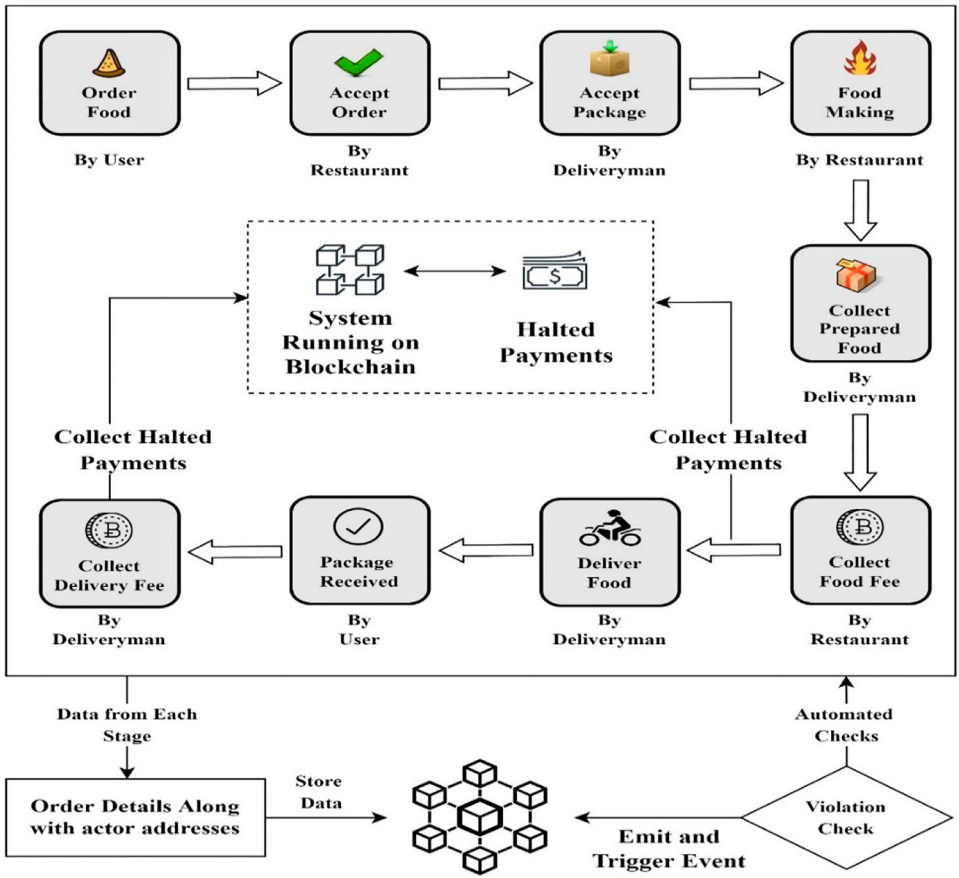

**Figure 3.** Detailed overview of proposed system.

***Decentralization***: The current integrated food aggregators application charges a percentage of commission on each request from both the restaurant and deliverymen; Due to the high cost of commissions and the fact that customers typically pay a lot when they use an app, the company's ability to expand is severely limited.

The delivery administration application can be withdrawn off the chain adopting blockchain technology, the owner of the external application can be replaced by the proprietor of the eatery, and customers' commissions can be minimized. At the same time, a more significant portion of the payment is distributed to both delivery drivers and restaurants. The problem-specific blockchain solution is as follows.

(1) **Smart contracts regulate commission rates:** A smart contract agreement can develop credibility between service providers and restaurants concerning food discounts and commission rates. It has the potential to protect restaurants from imposing excessive commission levels and being prey to greed, taking excess orders which cannot be served within the time limit. The smart contracts are configured so that if any of the restaurants violate the regulations, a portion of their commission is fined.

(2) **Confidence in the capabilities of delivery partners and service providers:** We can develop confidence between delivery partners and service providers by using blockchain-based contracts since Blockchain keeps track of all payments and makes sure none of them are compromised. Similar to restaurants, delivery persons also tend to accept many orders at once so that they get more commission within the same timeframe. Similar to restaurants, the delivery persons are also negatively rewarded for degrading service.

(3) **Remuneration by cryptocurrencies:** Our system incorporates cryptocurrency while completing a payment; customers will receive discounts and special offers when they pay for foods with cryptocurrencies.

(4) **Flawless delivery:** A customer's delivery might be inconvenient at times due to the deliveryman. If the deliveryman is late, he will be penalized. A portion of the delivery price will be deducted based on the time.

*Security*: Smart contracts safeguard every transaction. A smart contract is a computerized code that regulates data exchanges within a specified scope. Nobody, not even the administrator, can tamper with the data stored in Blockchain, smart contract-managed. Ledgers and established standards are used to deploy smart contracts. For these aforementioned reasons, blockchain-based systems are more secure than standard database-based applications.

Our system design is illustrated in Figure 3. Figure 3 demonstrates that the system's major actors are restaurants, deliverymen, and customers. Initially, the competent authority publishes the contract on the blockchain system. The steps start from food orders and end at collecting food delivery fees, and Blockchain holds some critical data. These steps are chronologically exhibited in Figure 3. Automated checks are done by the smart contracts on the different actors' data inside the system for security, traceability, and quality maintenance purpose.

### 3.2. Design of the system

The system is composed of actors such as food eateries, delivery personnel, and customers. Moreover, the contract deployer is another actor in this case. There a lot of elements will be used to create the mechanism of interaction between the actors and the system. The sections below illustrate the role of each actor and component.

*Actors:* While many different parties are involved, they are all threatened by the same thing: an unbreakable system that is not subject to change. Our suggested approach establishes connections between four actors through technical resources. The following section discusses the qualities of the actors.

(1) **Contract owner:** The contract owner wields a disproportionate amount of authority within the system. The owner is in charge of implementing the contract into operation on the system and ensuring that the rules are followed.

(2) **Restaurants:** Cafes, hotels, and many types of food preparation establishments are covered here. Their major objective is to cook food within a certain time frame and serve it to customers.
(3) **Deliverymen:** Food parcels are collected from restaurants and delivered to customers by the deliveryman. They employ their modes of transportation to ensure food is delivered on time.
(4) **Customer:** Customers are the vast majority of individuals who rely on deliverymen to bring their foods on time and who contribute significantly to the system by constantly increasing demand.

The primary objective is to develop a system in which these actors cooperate in terms of enhancing the transparency of these items.

***Components:*** Our system comprises many components to implement our system that has features such as immutability, availability, security, preventing third-party intermediaries, and automating. Blockchain technology and smart contracts are the principal components of our proposed system. The components and how they couple within our system are further explained in detail below.

***Blockchain:*** Blockchain technology contributes to the system's legitimacy and stability. One of the primary goals of this effort is to enhance the openness of restaurant food trackable data, tamper-proofing historical data and abolish commission business within the food industry. Several events occur at essential phases of the delivery process, and the data is recorded in the Blockchain's transaction log. This data can never be edited or tampered with without endangering the Blockchain. Our blockchain implementation has been done using Ethereum blockchain. Figure 4 demonstrates how our actors and the system interact in collaboration with the blockchain network.

At the same time, smart contracts automate our functionalities and eliminate the need for intermediary third parties (market controlling applications) through the use of a mix of properties, functions, events, and modifiers. Where attributes denote storage variables, functions denote a particular task execution, events denote the occurrence of a chosen set of statements, and modifiers indicate actor authority over the system. Our system is composed of three smart contracts developed in the Solidity programming language and is based on the Ethereum blockchain technology which we later merged into one. The first contract is for the customers. This contract holds the attributes, functions and modifiers related to a customer. The contract is demonstrated in Figure 5.

The second agreement covers restaurants and the third agreement covers deliveryman-related functionalities. The customer may place an order with any restaurant and wait for the deliveryman to come on time. Customers can track the status of their orders and receive notifications on the move. Figure 6 depicts the variables, functions, events, and modifiers that restaurant smart contracts contain, acquire, and execute. The restaurant contract (Figure 6) automates the gathering and keeping a record of customer data and placing it on the Blockchain. Self-testing is built into the system for rule violation from the smart contract and comparing it to the contract's ideal settings, the system takes the further action. As a result, the contract functions automatically on the Blockchain, ensuring verifiable information for future purchasers and quality control. The dates, as well as the amount and quantity sold, will all be recorded in this contract, at each stage from restaurant to customers. The second contract is used to track delivery-man once they have received the food for delivery which is shown in Figure 7.

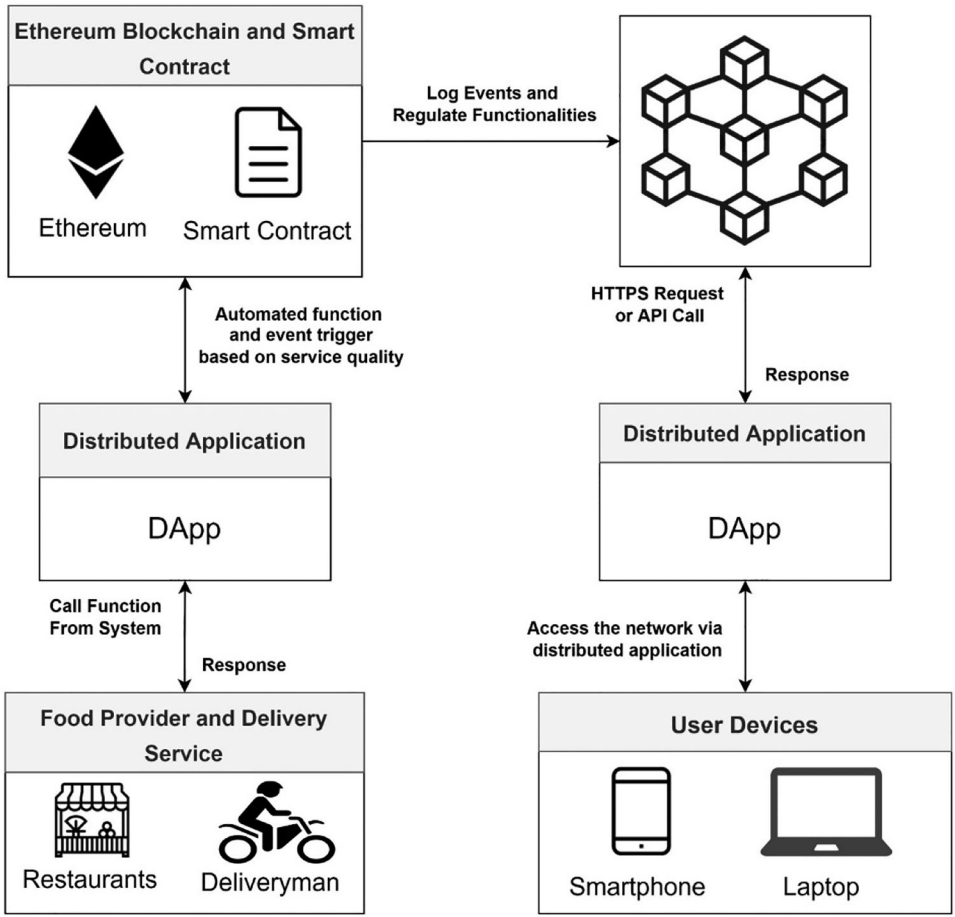

**Figure 4.** System-blockchain interaction.

One of the primary issues that our solution seeks to address is the legitimacy of order trackability data and making it auditable to general customers. Additionally, the second contract may be utilized to monitor deliveryman positions and their time for delivery and their quality of service (time-keeping) will affect their share of money in an order.

## 4. Implementation

We implemented and tested the blockchain mechanism in an Ethereum-based blockchain environment. We chose Solidity to write smart contracts. The Remix environment was used to develop and test the proposed system's fundamental prototype implementation. Our focus is not on the development of the entire system. However, our research demonstrates an architecture-driven method. We will cover the implementation in detail in this section and also present the testing result and simulations.

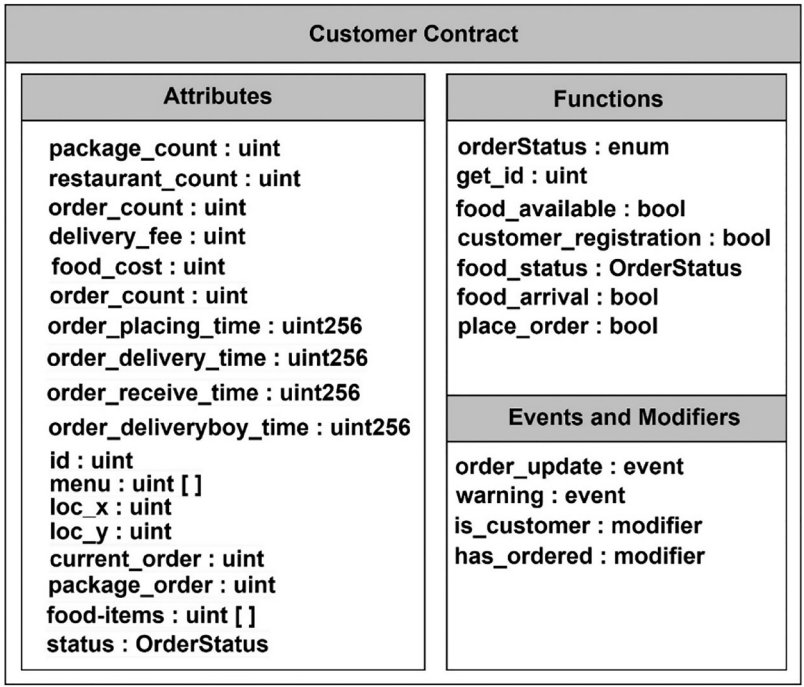

**Figure 5.** Smart contracts used for customers in the system.

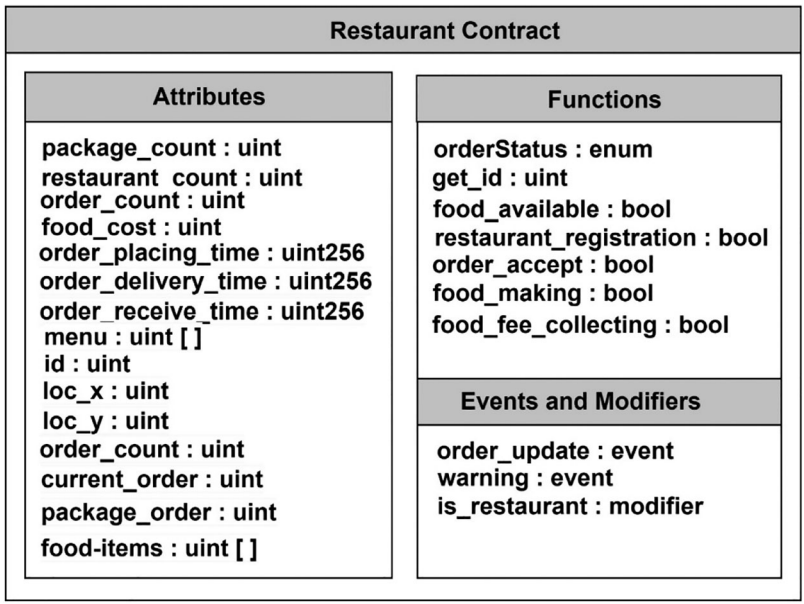

**Figure 6.** Restaurant's smart contracts in the system.

Our work demonstrates the usability and potential of blockchain technology in the realm of food delivery systems that ensure that all the actors involved do their job with honesty and the customers are satisfied. The core objective of this research is to depict how blockchain technology and smart contracts can be applied to automate the process of ordering food from a restaurant and having the food delivered to the customer's house by a deliveryman. Additionally, the system monitors the process's timing of food delivery. The connections between system components and actors are evident in Figures 3, 4 and 8. We divided the process into four segments: the registration phase for all actors, the customer to restaurant phase, the restaurant to deliveryman phase, and the deliveryman to customer phase. The operation of each segment is detailed below.

**Figure 7.** Smart contracts used for deliveryman in the system.

### 4.1. Registration

All three actors involved within our system, the customer, the restaurant, and the deliveryman's package, had to first registered to the system. All actors have an identical process. During the registration phase, our system primarily stores the addresses of customers, restaurants, and packages. Additionally, as previously stated, all three actors use a similar registration process, but the restaurant registration algorithm is slightly different. The distinction is that during restaurant registration, the restaurants need to provide their menu to the system where the menu is an array containing the food's id.

### 4.2. Customer to restaurant

Once a customer places an order in the system, the smart contract first checks if the food item is available or not via the food_available() method. Given that the item is available, the customer's order for food is placed to the network by executing the place_order() method. An ordered hash is generated for each order. Following that, the restaurant will accept the order by using the accept_order() method.

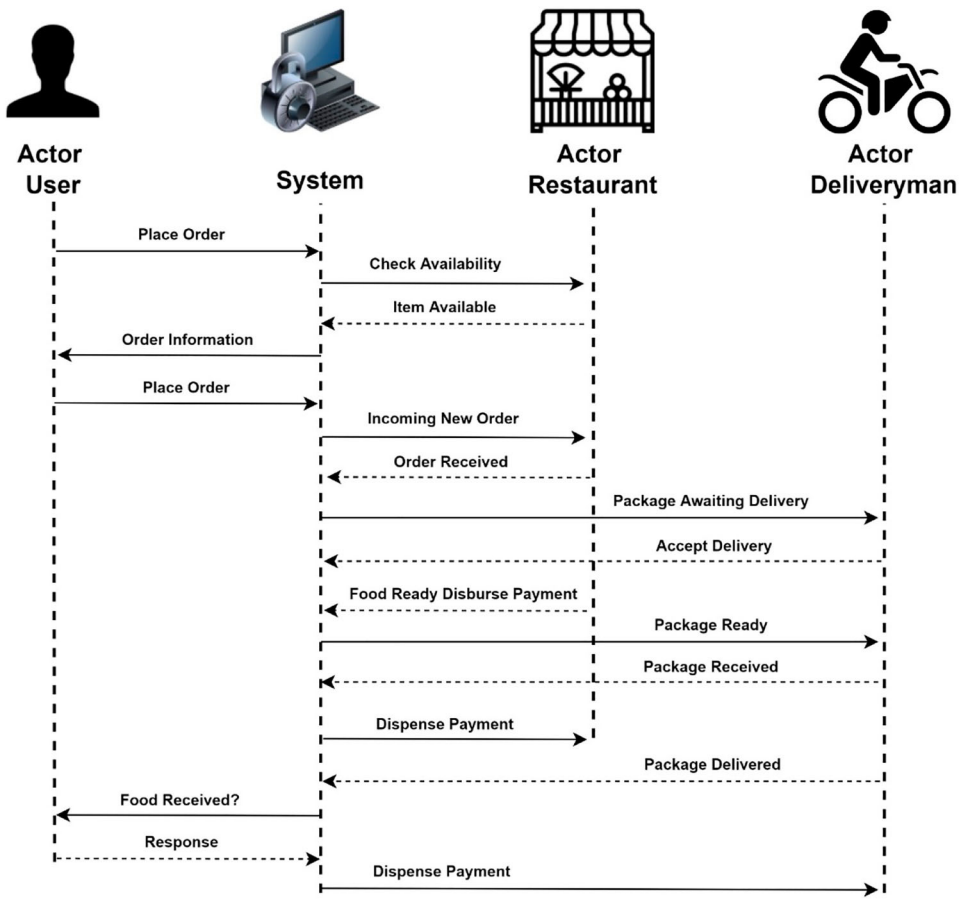

**Figure 8.** Sequence flow diagram for one order.

### 4.3. Restaurant to deliveryman

Our third phase is the connection between the restaurant and the deliveryman. In our second phase, the customer places an order and the restaurant accepts it. Additionally, the Customer to Restaurant process generated an ordered hash. In this stage, the deliveryman will get a notification showing that restaurant 'X' has a package waiting to be delivered. Accepting the order from the restaurant, the deliveryman will accept the food delivery which invokes the accept_package() method within the system with the order hash being designated to the deliveryman's address. Following that, the restaurant will start preparing the food by executing the food_making() method and passing the order hash as an argument for this method. After completing the meal, the deliveryman will receive it as a package by executing the collect_food() method in the background. Finally, the restaurant can only collect the total amount for food by invoking the method food_fee_collecting() inside the system. Additionally, we incorporate the concept of food preparation in a predetermined time frame and maximum order constraint during the Restaurant to Deliveryman connection phase. If a time violation occurs, or if the restaurant takes an excessive number of orders at a time, making delay while handing the parcel over to the deliveryman, the restaurant's food preparation payment will be reduced by a certain percentage which will increase/decrease based on the restaurant's quality control. The algorithm for determining the set time violation is depicted in Algorithm 1.

**Algorithm 1: Time violation and restaurant payment algorithm in phase three**

```
            Result: 1. Smart Contract emits time violation.    2. Constraint-based restaurant
                    payment.
01          food_making():
              order placement time : block.timestamp while restaurant starts preparing food
02          collect_food():
              order delivery time : block.timestamp while delivering the food.
            Food Fee Collection by the Restaurant
03          food_fee_collecting():
04              if (delivery time – placement time) is greater than the time
                promised to customer then
05                      triggering warning event;
06                      emit warning event in Blockchain;
07                      reduce amount by 'x' percentage;
08              else
09                      restaurant gets full payment;
10              end
11          end
```

### 4.4. Deliveryman to customer

The final phase of our proposed system is the interaction between the Deliveryman and the Customer. After the food is ready, the deliveryman will receive the package from the restaurant, and the deliveryman is responsible for delivering the parcel to the customer within a certain time. The deliveryman application will invoke the deliver_food() method in the system, which will create a parcel information block within the Blockchain with the same hash that was generated while the user created the food request. After that, the food_arrival() method gets invoked once the customer confirms food arrival. Finally, only after the customer confirmation, the deliveryman will be able to collect the delivery fees by invoking the collect_delivery_fee() method in the background. Additionally, similar to our third phase, the Restaurant to Deliveryman connection phase, our system implements the constraints of food delivery in a predetermined time frame and a maximum number of orders received by the deliveryman in this phase, which is also our system's final phase. If a specified time violation occurs, or if the deliveryman takes an excessive amount of time delivering the food or handing it over to the customer, the deliveryman's payment will be reduced as a penalty. No one likes their food cold and past its prime. Paying extra for having the food in the comfort of home comes with an expectation of having the food as it just came straight out from the kitchen. But deliverymen tend to accept arrays of orders to deliver them together which impairs both qualities of service and food. Algorithm 2 shows the deliveryman constraints and fee collection process. At the conclusion of phase four, our entire system had successfully completed the food delivery process from restaurant to customer. For a clear understanding, Figure 8 depicts the system's sequence diagram, which is a graphical flow representation of all actors interacting in a food order and delivery process with our system.

**Algorithm 2: time violation and deliveryman payment algorithm in phase four**

```
              Result: 1. Smart Contract emits time violation.
                      2. Constraint-based deliveryman payment.
01            collect_food():
                 delivery start time : block.timestamp while deliveryman
                 receives the parcel.
02            food_arrival():
                 order arrival time : block.timestamp while
                 customer receives the food.
              Food Fee Collection by the Restaurant
03              delivery_fee_collecting():
04                 if (arrival time–delivery start time) is greater than the time
                   promised to customer then
05                    triggering warning event;
06                    emit warning event in Blockchain;
07                    reduce amount by 'x' percentage;
08                 else
09                    deliveryman gets full payment;
10                 end
11              end
```

## 5. Testing and analysis

### 5.1. Testing

Our team designed and implemented the system architecture, which included the interface between smart contracts and the Ethereum blockchain. In addition to providing an Ethereum wallet, accounts that have been pre-loaded with dummy ether cryptocurrency, and an environment for building smart contracts in the Solidity programming language, Remix also provides a web-based integrated development environment (IDE). Additionally, Remix offers us with the appropriate infrastructure for executing and deploying the contract on the Ethereum blockchain, which fulfils our requirements and objectives. We will demonstrate our contracts communicating with the Blockchain in this part and the process by which they interact.

The three actors within our system communicated within themselves to execute different functionalities and, in the process make various requests, responses and checks. Rest of this section will depict a food order by the customer successfully creating a block inside the blockchain network and the progression will conclude with the customer receiving the order. During the food making and delivery phase, certain violations may occur which will end up in penalty for that user which has also been shown. After all the actors are registered in our system, the second phase can commence which is the interaction between the customer and the restaurant. During this step, the customer places an order for food using the place_order() method. As depicted in Figure 9, the logs contain all the food order process information. The Blockchain logs the food items or food id, the restaurant id, the order id, and the price along with customer and restaurant hashes.

The third and last phase of the system ensures that food reaches the customer's hand from the restaurant. In these two stages, first, the restaurant accepts the order stating its

```
hash           0xdb566cc9966a613dc678f7e06c7a4c655604407a10e60b98b1ba8f045f3070bf
input          0x2e7...00002
decoded input  {
                   "uint256[] food_items": [
                       "1",
                       "2"
                   ],
                   "uint256 restaurant_id": "1"
               }
decoded output {
                   "0": "bool: true"
               }
logs           [
                   {
                       "from": "0xd9145CCE52D386f254917e481eB44e9943F39138",
                       "topic": "0x0c2f387cf39fbb79382e223d10c940a8f2eaada67a10a1d377ff2e6f54b450d0",
                       "event": "order_update",
                       "args": {
                           "0": "1",
                           "1": 0,
                           "order_id": "1",
                           "status": 0
                       }
                   }
               ]
```

**Figure 9.** A Successful order placed inside the blockchain.

availability and the based on the food type, the restaurant will ask for a certain amount of time to prepare the food. A food request received by the restaurant has been shown in Figure 10 where relevant information like customer and restaurant address, order details and food making time is logged inside the Blockchain.

An order placed in a restaurant also notifies the deliveryman and someone accepts the delivery request. After the restaurant has prepared the food, the deliveryman picks up the food to deliver it to the customer. The delivery information such as deliveryman and customer hash, the food package id (hash) and the delivery time is logged into the network. Figure 11 shows a successful delivery by the restaurant received by the deliveryman.

The food package's location and time are tracked along the way till it reaches to the customer. Figure 12 shows the blockchain logs of an order received by the customer along with the associated address of the customer, package, and deliveryman.

We previously discussed the time violation event, referred to as a warning() event in the system, in which restaurants who deliver food late would have a portion of their payment removed as a penalty while implementing the food_fee_collecting() method. Similarly, if the deliveryman delivers the food late, they will also be charged a penalty while implementing the and collect_delivery_fee() method. These two cases are shown in Figure 13, where 13(a) shows the violation log for the restaurant and 13(b) illustrates the same scenario for a deliveryman.

## 5.2. Analysis

This part aims to analyze the proposed system and identify the motivations why Blockchain-based food delivery should be integrated.

### 5.2.1. Advantages

Blockchain-Powered food delivery service aimed at resolving industry-wide trust issues. Referral and loyalty reward programs entice users to stay on the site. Liquidity is

```
hash             0xd9a9cb592150cecd46453561f3714dde4820f1a53b160ae2d551a7e415529a97
input            0xfdd...00001
decoded input    {
                     "uint256 order_id": "1"
                 }
decoded output   {
                     "0": "bool: true"
                 }
logs             [
                     {
                         "from": "0xd8b934580fcE35a11B58C6D73aDeE468a2833fa8",
                         "topic": "0x0c2f387cf39fbb79382e223d10c940a8f2eaada67a10a1d377ff2e6f54b450d0",
                         "event": "order_update",
                         "args": {
                             "0": "1",
                             "1": 1,
                             "order_id": "1",
                             "status": 1
                         }
                     },
                     {
                         "from": "0xd8b934580fcE35a11B58C6D73aDeE468a2833fa8",
                         "topic": "0x0571c17fabdbebea862518abb447980c8b2af18b117ce124e8eb3c75a596112d",
                         "event": "event1",
                         "args": {
                             "0": "Your order has been placed",
                             "msg": "Your order has been placed"
                         }
                     }
                 ]
```

**Figure 10.** A successful receive of order by the restaurant.

ensured by cryptocurrency usage. One way to address discernibility difficulties and assure straightforwardness, one solution is to deploy block tie technology to record information from the consecutive order in a way that makes subsequent control difficult. There are options such as postponing operational events, trading hard currency at each counter, and reducing functional proficiency. Inadequate data security. Additional personnel behind the counters. Despite the fact that there is currently a lack of clarity about the taxation of cryptographic forms of money. Table 1 illustrates

```
hash             0x6c2623193ec28d0bc47fdd3bacf838475a21aeef94ad4d88ed08c38394930cff
input            0x36d...00001
decoded input    {
                     "uint256 order_id": "1"
                 }
decoded output   {
                     "0": "bool: true"
                 }
logs             [
                     {
                         "from": "0xEF9f1ACE83dfbB8f559Da621f4aEA72C6EB10eBf",
                         "topic": "0x0c2f387cf39fbb79382e223d10c940a8f2eaada67a10a1d377ff2e6f54b450d0",
                         "event": "order_update",
                         "args": {
                             "0": "1",
                             "1": 4,
                             "order_id": "1",
                             "status": 4
                         }
                     },
                     {
                         "from": "0xEF9f1ACE83dfbB8f559Da621f4aEA72C6EB10eBf",
                         "topic": "0xd61481a11e5449d64f97e1edf49aec06a9ff79cef718c30115a331a73ae49cb4",
                         "event": "event2",
                         "args": {
                             "0": "Your package has been received by the deliveryman",
                             "msg": "Your package has been received by the deliveryman"
                         }
                     }
                 ]
```

**Figure 11.** Food package received by the deliveryman.

**Figure 12.** Food package received by the customer.

**Figure 13.** Food package received by the customer.

a Comparative examination of our system with previous research employing Blockchain and smart contracts difficulties and advantages exemplify our system's analysis. The following are the findings from the comparison of the system to existing methods indicated above in Table 1.

**Table 1.** Comparing the system to existing method.

| Outcome | Shyamala Devi et al. (2019) | Rani and Vishali (2021) | Ngamsuriyaroj et al. (2018) | Bartolini et al. (2020) | Proposed system |
| --- | --- | --- | --- | --- | --- |
| Traceability | ✗ | ✓ | ✗ | ✗ | ✓ |
| Integrity | ✗ | ✗ | ✓ | ✓ | ✓ |
| Smart contracts | ✓ | ✓ | ✓ | ✓ | ✓ |
| Time violation | ✗ | ✗ | ✗ | ✗ | ✓ |
| Penalty | ✗ | ✗ | ✗ | ✗ | ✓ |
| Eliminate third-party | ✗ | ✓ | ✗ | ✗ | ✓ |

- To encourage greater adoption of Blockchain, our suggested system incorporated penalties for time violations. Both the time violation and the penalty are unique to each of the above methods mentioned.
- Immutable data storage and eliminating third-party intermediaries are not typical features in all of these methods, but they are in our suggested system, encouraging greater adoption.

### 5.2.2. Challenges

Though our system tries to prevent faults and blunders in online food delivery systems with the use of advanced regulatory contracts and immutable and verifiable ledger technology, there are still some systemic issues to resolve. Blockchain technology is still in its early stages, finding it challenging to adapt it for real-world applications. Still being execution-costly, mining time costly, and on the other hand, the distributed network is computationally expensive while creating a trade-off between reliability and ease of use [original 7]. Editing or modifying data on the Blockchain is nearly impossible, making it impossible to correct errors. While immutability is a core characteristic of Blockchain, it differs from typical database systems; Blockchain does not enable us to edit the data associated with a validated transaction owing to its reliable security methods. Smart contracts are installed to take charge of the system and make the procedures easier to do. Once it is placed on the Blockchain, it cannot be altered, and the features within it are not open to addition or modification. The initial configuration of the blockchain ecosystem requires a modest sum of money. There's also the question of the legality of trading via cryptocurrency based on country laws. Gas costs to run and execute functions are shown in Table 2.

**Table 2.** System costs for execution and transaction.

| Task | Actor | Transaction cost | Execution cost |
| --- | --- | --- | --- |
| registration() | Restaurant | 385,355 | 385,355 |
| | Customer | 93,566 | 93,566 |
| | Package | 140,108 | 140,108 |
| place_order() | Customer | 223,985 | 223,985 |
| order_accept() | Restaurant | 51,981 | 51,981 |
| accept_package() | Deliveryman | 75,571 | 75,571 |
| food_making() | Restaurant | 55,157 | 55,157 |
| collect_food() | Deliveryman | 57,026 | 57,026 |
| food_fee_collecting() | Restaurant | 57,133 | 57,133 |
| deliver_food() | Deliveryman | 28,677 | 28,677 |
| food_arrival() | Customer | 77,356 | 77,356 |
| collect_delivery_fee() | Deliveryman | 333,44 | 333,44 |

# 6. Conclusion

Blockchain's role in the food delivery industry was the main focus of this research and how its use can ensure data's accessibility, privacy, integrity which also prevents data tampering and to build trust between restaurants, customers, and deliveryman. Using a penalty system for food delivery on the blockchain simplified things for restaurants and consumers alike, as they will be able to produce, deliver, and receive food on time. Customers can safely assume that they will receive their food on time and that the restaurant and deliveryman will receive payment as well if they prepare and deliver the food on time. A commission-based market eventually jeopardizes the rights and advantages from the customer end while the other actors making a fortune out of the system flaws. To put an end to these flaws, our study implemented a blockchain and smart contract-based system and our testing result shows that the implementation tackles the issues tried to be solved. The entire system is made secure, as blockchain technology ensures its robust security. The implementation of our study has been made flexible and broad in reach. Any process may be traced, developed, and enhanced using the same methodology. In future work, installing the system on a permissioned blockchain and innovating features as customers can also get the penalty, checking food quality before it is even prepared, etc. is our long-term goal is our long-term goal.

## Disclosure statement

No potential conflict of interest was reported by the author(s).

## Funding

This work was supported by the Faculty Research Grant [CTRG-21-SEPS-19], North South University, Bashundhara, Dhaka 1229, Bangladesh.

## Notes on contributors

*A. A. Talha Talukder* is currently studying B.Sc. in the department of Electrical & Computer Engineering, North South University. He is the COO and the founder of a startup named 'Instantkaj.com' which is for freelancers in Bangladesh. He is also working part-time software developer at an IT company named Anza Corporation Ltd. He along with his team secured the first runner-up position is the ACM-Innovation Challenge held by the ECE department of North South University for their undergraduate final year design project. His main research interests include data analytics, computer vision, machine learning and blockchain.

*Md. Anisul Islam Mahmud* is now pursuing BSc. in Electrical and Computer Engineering at North South University. He is working as an undergraduate teaching assistant at North South University. He developed a Bengali language-based chatbot for university students. He and his team secured the first runner-up position is ACM-Innovation Challenge held by the ECE department of North South University for their undergraduate final year design project. His main research interests are in blockchain, machine learning, deep learning, image processing and computer vision.

*Arbiya Sultana* is now pursuing B.Sc. in Electrical and Computer Engineering at North South University. She has interest in user interface design and front-end of software architecture. She worked on many software projects. Apart from her interest in software, she also has great passion for research and development in the field of machine learning and blockchain. She and her team secured the first runner-up position is ACM-Innovation Challenge held by the ECE department of North South

University for their undergraduate final year design project. Her research interests are in blockchain and machine learning.

*Tahmid Hasan Pranto* received the bachelor's degree in computer science and engineering from North South University, Dhaka, Bangladesh. He is currently working as a Research Assistant at the ECE Department, North South University, under the supervision of Prof. Dr. Rashedur M. Rahman. He has keen interest in emerging technologies like blockchain and machine learning. He has published research works in peer-reviewed journals like Computer Science (PeerJ), Applied Artificial Intelligence, IEEE Access, and Cybernetics and Systems. He has also published in conference proceedings like IWANN, 2021. His current research explores the scope and feasibility of artificial intelligence in decentralized platforms by employing incremental machine learning and federated learning approaches in the blockchain-enabled decentralized systems.

*AKM Bahalul Haque* is currently a Junior Researcher with the Department of Software Engineering, LUT University. Earlier, he was a Lecturer at the Department of Electrical and Computer Engineering, North South University. His works have been accepted and published in international conferences and peer-reviewed journals, including IEEE Access, Expert Systems, Cybernetics and Systems, various international conference proceedings, Taylor and Francis Books, and Springer Book. His research interests include explainable AI, blockchain, data privacy and protection, and human-computer interaction.

*Rashedur M. Rahman* (Senior Member, IEEE) received the M.S. degree from the University of Manitoba, Winnipeg, MB, Canada, in 2003, and the Ph.D. degree from the University of Calgary, Canada, in 2007. He currently works as a Professor in the Electrical and Computer Engineering Department at North South University, Dhaka, Bangladesh. He published more than 200 research articles in the area of parallel and distributed computing, cloud and grid computing, data and knowledge engineering. His current research interests include cloud load characterization, VM consolidation, and the application of data mining and fuzzy logic in different decision-making problems. He is also on the editorial committee of many international journals.

## ORCID

*Rashedur M. Rahman* http://orcid.org/0000-0002-4514-6279